\pdfoutput=1

\documentclass[11pt,twoside,a4paper,cmspaper,final,collab]{cms-tdr}
\def\svnVersion{274090}\def\cmsCernNo{2013-037}\def\cmsCernDate{\today}\def\cmsMessage{Published in the Journal of High Energy Physics as \href{http://dx.doi.org/10.1007/JHEP01(2015)063}{\doi{10.1007/JHEP01(2015)063.}}}

\begin{document}\cmsNoteHeader{BPH-12-011}

\hyphenation{had-ron-i-za-tion}
\hyphenation{cal-or-i-me-ter}
\hyphenation{de-vices}
\RCS$Revision: 274028 $
\RCS$HeadURL: svn+ssh://svn.cern.ch/reps/tdr2/papers/BPH-12-011/trunk/BPH-12-011.tex $
\RCS$Id: BPH-12-011.tex 274028 2015-01-16 09:31:16Z malvezzi $
\providecommand{\PBc}{\ensuremath{\cmsSymbolFace{B}_\cmsSymbolFace{c}}\xspace}
\providecommand{\PBpc}{\ensuremath{\cmsSymbolFace{B}^+_\cmsSymbolFace{c}}\xspace}
\providecommand{\PBmc}{\ensuremath{\cmsSymbolFace{B}^-_\cmsSymbolFace{c}}\xspace}
\providecommand{\PBpmc}{\ensuremath{\cmsSymbolFace{B}^\pm_\cmsSymbolFace{c}}\xspace}
\providecommand{\PX}{\ensuremath{\cmsSymbolFace{X}}\xspace}
\newcommand{\BcJpsiPi}{\ensuremath{\PBpc \to \JPsi \Pgpp}\xspace}
\newcommand{\BcJpsiPiTitle}{\ensuremath{\PBpmc \to \JPsi \Pgppm}\xspace}
\newcommand{\BJpsiK}{\ensuremath{\PBp \to \JPsi \PKp}\xspace}
\newcommand{\BJpsiKTitle}{\ensuremath{\PBpm \to \JPsi \PKpm}\xspace}
\newcommand{\BcJpsiK}{\ensuremath{\PBpc \to \JPsi \PKp}\xspace}
\newcommand{\BcJpsiPiPiPi}{\ensuremath{\PBpc \to \JPsi \Pgpp \Pgpp \Pgpm }\xspace}
\newcommand{\BcJpsiPiPiPiTitle}{\ensuremath{\PBpmc \to \JPsi \Pgppm \Pgppm \pi^{\mp} }\xspace}
\newcommand{\BcJpsiPiPiPiPiO}{\ensuremath{\PBpc \to \JPsi \Pgpp \Pgpp \Pgpm \Pgpz}\xspace}
\newcommand{\BcJpsiPiPiO}{\ensuremath{\PBpc \to \JPsi \Pgpp \Pgpz}\xspace}
\newcommand{\BcJpsiKKPi}{\ensuremath{\PBpc \to \JPsi \PKp \PKm \Pgpp }\xspace}
\newcommand{\JpsiPi}{\ensuremath{\JPsi \Pgpp}\xspace}
\newcommand{\JpsiK}{\ensuremath{\JPsi \PKp }\xspace}
\newcommand{\JpsiPiPiPi}{ \ensuremath{\JPsi \Pgpp \Pgpp \Pgpm}\xspace}
\newcommand{\RatioBcBp}{\ensuremath{R_{\cPqc/\cPqu}}\xspace}
\newcommand{\RatioBcBc}{\ensuremath{R_{\PBc}}\xspace}
\newcommand{\RatioBcBpDef}{\ensuremath{R_{\cPqc/\cPqu} = \big({\sigma(\PBpc)\mathcal{B}(\PBpc \to \JPsi \Pgpp)}\big)/ \big({\sigma(\PBp)\mathcal{B}(\PBp \to \JPsi\PKp)}\big)}\xspace}
\newcommand{\RatioBcBpDeffirst}{\ensuremath{R_{\cPqc/\cPqu} \equiv \big({\sigma(\PBpc) \mathcal{B}(\PBpc \to \JPsi \Pgpp)}\big)/ \big({\sigma(\PBp) \mathcal{B}(\PBp \to \JPsi\PKp)}\big)}\xspace}
\newcommand{\RatioBcBcDeffirst}{\ensuremath{R_{\PBc} \equiv {\mathcal{B}(\PBpc \to \JPsi \Pgpp\Pgpp\Pgpm)}/{\mathcal{B}(\PBpc \to \JPsi \Pgpp)}}\xspace}
\newcommand{\RatioBcBcDef}{\ensuremath{R_{\PBpc} =  {\mathcal{B}(\PBpc \to \JPsi \Pgpp\Pgpp\Pgpm)}/{\mathcal{B}(\PBpc \to \JPsi \Pgpp)}}\xspace}

\cmsNoteHeader{BPH-12-011} % This is over-written in the CMS environment: useful as preprint no. for export versions
\title{Measurement of the ratio of the production cross sections times branching fractions of
\BcJpsiPiTitle and \BJpsiKTitle and
$\mathcal{B}(\BcJpsiPiPiPiTitle)/\mathcal{B}(\BcJpsiPiTitle)$
in pp collisions at $\sqrt{s} = 7\TeV$}

\date{\today}

\abstract{
The ratio of the production cross sections times branching fractions
$\big({\sigma(\PBpmc) \mathcal{B}(\PBpmc \to \JPsi \Pgppm)}\big)/ \big({\sigma(\PBpm) \mathcal{B}(\PBpm \to \JPsi\PKpm)}\big)$
is studied in proton-proton collisions at a center-of-mass energy of 7\TeV with the CMS detector at the LHC.
The kinematic region investigated requires $\PBpmc$ and $\PBpm$ mesons with transverse momentum $\pt > 15$\GeV and
rapidity $\abs{y} < 1.6$. The data sample corresponds to an integrated luminosity of  5.1\fbinv. The ratio is determined to be
$[0.48 \pm 0.05\stat\pm 0.03\syst\pm 0.05\,(\tau_{\PBc})]\%$. The $\PBpmc \to \JPsi \Pgppm \Pgppm \pi^\mp$ decay is also observed in the same data sample. Using a model-independent method developed to measure the efficiency given the presence of
resonant behaviour in the three-pion system, the ratio of the branching fractions $\mathcal{B}(\PBpmc \to \JPsi \Pgppm \Pgppm \pi^\mp)/\mathcal{B}(\PBpmc \to \JPsi \Pgppm)$ is measured to be
$2.55 \pm 0.80\stat\pm 0.33\syst^{+ 0.04}_{-0.01}\,(\tau_{\PBc})$, consistent with the previous LHCb result.
}

\hypersetup{%
pdfauthor={CMS Collaboration},%
pdftitle={Measurement of the ratio of the production cross sections times branching fractions of Bc+/- to J/psi pi+/-
and B+/- to J/psi K+/- and  B(Bc+/- to J/psi pi+/- pi+/- pi-/+)/B(Bc+/- to J/psi pi+/-) in pp collisions at sqrt(s) = 7 TeV},%
pdfsubject={CMS},%
pdfkeywords={CMS, physics, B physics}}

\maketitle

\section{Introduction}

The pseudoscalar $\PBpc\,(\PBmc)$ meson, the ground state of the $\cPaqb\cPqc\,(\cPqb\cPaqc)$ system, is the lightest particle containing two heavy quarks
of different flavors and thus represents a unique laboratory in which to study heavy-quark dynamics.
The investigation of the \PBpc meson properties (charge conjugation is implied throughout this paper) is of special interest compared to the flavor symmetric heavy-quarkonium (\bbbar, \ccbar) states, and provides a new testing ground for predictions in the context of effective models inspired by quantum chromodynamics \cite{theo2}.
The decay processes of the
\PBpc meson can be generically divided into three classes: those involving the decay of the b quark, the decay of the c quark, and the annihilation of the b and c quarks \cite{theo2,ann1,ann2}.
The $\cPqb \to \cPqc$ transition, accounting for about 20\% of the decay rate \cite{theo2}, offers an easily accessible experimental signature, having a high probability of producing a \JPsi meson.
Consequently, the first \PBpc observation was made in the semileptonic channel
$\PBpc \to \JPsi \ell^{+} \nu \, (\ell=\Pe,\mu)$ by the CDF Collaboration \cite{CDF_sl}.

Information on the production and decays of the $\PBpc$ meson and its possible excited states is quite limited. Production characteristics have been studied at the Tevatron for $\pt> 4\GeV$  and $\abs{y}< 1$ and by LHCb
in the kinematic region  $\pt> 4\GeV$, $ 2.5<\eta <4.5$.

The advent of the CERN LHC has opened a new era for \PBpc investigations; a rich program of measurements involving new decay modes is being carried out by the LHCb Collaboration \cite{lhcb_kkpi,
lhcb_bctobspi, lhcb_jpsids, lhcb_psipi, lhcb_JpsiK, lhcb3to1, lhcb_3pi2pi}. The ATLAS experiment has recently observed a new state whose mass is consistent with the predicted mass for the second S-wave state of the \PBpc meson \cite{atlasBcStar}.
The CMS experiment, owing to its excellent muon identification system and tracking detectors, is particularly well suited to the study of final states containing \JPsi mesons, where $\JPsi \to \Pgmp\Pgmm$.

In this Letter, the $\PBpc$ production cross section relative to that of $\PBp$  at 7\TeV center of mass energy is presented in a kinematic region complementary to that accessible to LHCb.
This is done by comparing the \BcJpsiPi mode to \BJpsiK, which has a similar vertex topology.
The ratio of their production cross sections times branching fractions
\RatioBcBpDeffirst is measured for \PBpc transverse momentum $\pt>15\GeV$ and
in the central rapidity region, $\abs{y} < 1.6$.
This measurement contributes to a more complete understanding of \PBpc production in pp collisions.

We also observe the decay \BcJpsiPiPiPi, previously measured by LHCb \cite{lhcb3to1}.
The three-pion system exhibits resonant structure that is not well-determined, but can affect the reconstruction efficiency.
A model-independent efficiency correction method is developed and implemented in this analysis, which can be considered for future high-statistics analyses of multibody decays.
The ratio of the branching fractions \RatioBcBcDeffirst, which is independent of the poorly known \PBpc production cross section, is also presented; our measurement provides the first confirmation of the LHCb Collaboration's result \cite{lhcb3to1}.

\section{CMS detector}

The central feature of the CMS apparatus is a superconducting solenoid of 6\unit{m} internal diameter and 3.8\unit{T} field.
Within the superconducting solenoid volume are silicon pixel and strip trackers, a crystal electromagnetic calorimeter,
and a brass/scintillator hadron calorimeter. Muons are measured in gas-ionization detectors embedded in the steel return yoke.
The subdetectors relevant for this analysis are the silicon tracker and the muon systems.
The inner tracker measures charged particles within the pseudorapidity range $\abs{\eta}< 2.5$.
It consists of layers totaling 66 million 100$\times$150 $\mum^2$ silicon pixels and 9.6 million silicon strips with pitches ranging from 80 to 183\mum.
Muons are measured in the pseudorapidity range $\abs{\eta}< 2.4$ with detection planes constructed using three technologies: drift tubes,
cathode strip chambers, and resistive-plate chambers.
The first level of the CMS trigger system, composed of custom hardware processors, uses information from the calorimeters and muon
detectors to select the most interesting events in a fixed time interval of less than 4\mus.
The high level trigger processor farm further decreases the event rate from around 100\unit{kHz} to approximately 400\unit{Hz} before data storage.
A more detailed description of the CMS detector, together with a definition of the coordinate system used and the relevant kinematic variables, can be found in Ref.~\cite{CMSdet}.

\section{Event selection}
This analysis is based on pp data collected with the CMS detector at a center-of-mass energy of 7\TeV in 2011. Events selected with displaced-vertex dimuon triggers are considered, corresponding to an integrated luminosity of  5.1\fbinv.
The analysis is driven by the \JPsi meson reconstruction. The dimuon triggers apply topological and kinematic selections on dimuon candidates:
$\cos\alpha >0.9$, where $\alpha$ is the pointing angle, in the transverse plane, between the dimuon momentum and
the direction from the mean pp collision position (beam spot) to the dimuon vertex;
$L_{xy}/\sigma_{xy}>3$, where $L_{xy}$ is the transverse distance between the beam spot and the dimuon vertex, and
$\sigma_{xy}$ is the corresponding uncertainty.
In addition, the two muons must have opposite charges, dimuon $\pt > 6.9$\GeV, satisfy an invariant mass
requirement 2.9 $<m(\Pgmp\Pgmm)< 3.3$\GeV, and have a mutual distance of closest approach in the transverse plane of less than 0.5\unit{cm}.
Trigger selection requirements on the $\chi^2$ probability of the dimuon kinematic fit ($P_\mathrm{VTX}$) and muon transverse momentum $\pt(\mu)$ were made more restrictive as the luminosity increased and ranged from $P_\mathrm{VTX}> 0.5$\% to 15\% and
up to $\pt(\mu)>4\GeV$, respectively. A requirement on the muon pseudorapidity, $\abs{\eta(\mu)}< 2.2$, is also applied.

Monte Carlo (MC) simulations are employed to design the offline selection, assess the reconstruction efficiency, and study systematic effects.
The \PBpc signal events are simulated using a dedicated generator (\textsc{bcvegpy}) \cite{Bcveg1, Bcveg2} interfaced with the
\PYTHIA simulation program (version 6.424, Z2 tune \cite{pythia,z2tune}), which hadronizes the whole event. Unstable particle decays are simulated with
\EVTGEN \cite{evtgen} and the detector response with \GEANTfour \cite{geant}.

The offline selection starts from \JPsi candidates reconstructed from pairs of oppositely charged muons. Muons are identified through the standard CMS muon reconstruction procedure \cite{CMSMuon} and are required to have a track matched with at least one muon segment, a track fit $\chi^2$ per degree of freedom less than 1.8, at least 11 hits in the tracker with at least 2 from the pixel detector, and a transverse (longitudinal) impact parameter less than 3\unit{cm}\,(30\unit{cm}). Offline requirements on the dimuon pair are tightened, with respect to those of the trigger, requiring a dimuon $\pt >7.1\GeV$ and $L_{xy}/\sigma_{xy} \geq5$.

The \BcJpsiPi ($\mathrm{B^{+}}\to\JPsi\PKp$) candidates are formed by combining a \JPsi candidate with one track, assuming that it is a pion (kaon). The track must not be identified as a muon.
The \BcJpsiPiPiPi candidates are analogously formed by combining a \JPsi candidate with three tracks, assuming that they are pions
and requiring that the total charge is ${+}1$.
The pion (kaon) candidates are required to have a track fit $\chi^2$ less than three times the number of degrees of freedom;
$\ge$6 tracker hits; $\ge$2 pixel hits; $\abs{\eta}< 2.4$; and $\pt > 0.9$\GeV.
The three-dimensional impact parameter between each pion (kaon) and the $\JPsi$ vertex is required to be less than 6 times its uncertainty to reduce combinatorial background and the effect of the number of simultaneous pp interactions per bunch crossing (pileup).
The decay vertex is reconstructed using a kinematic vertex fit \cite{kinefit}, which constrains the invariant mass of the two muons
to the nominal \JPsi mass. After the vertex fit, the track parameters are re-estimated at the fitted vertex to improve their resolution.
To reduce backgrounds, only the highest-transverse-momentum \PBpc or \PBp candidate is retained per event.
This method has been studied using the MC samples and found to select the right candidate in  99.3\% (99.2\%) of the \BcJpsiPi (\BJpsiK) events and 91\% of the \BcJpsiPiPiPi events.
In the \BcJpsiPiPiPi data sample, this requirement reduces the background by more than a factor of four.

Additional topological requirements are made to improve the signal-to-background ratio, as discussed below.

\section{\texorpdfstring{\BcJpsiPi}{Bc to J/psi pi}  and \texorpdfstring{\BJpsiK}{B to J/psi K} decays}
\label{sec:BcBpSignals}
The selection criteria for the \BcJpsiPi decay have been optimized in the kinematic region $\pt >15$\GeV and $\abs{y}< 1.6$ by maximizing $S/\sqrt{(S+B)}$ as a figure of merit,
where $S$ is the signal yield obtained from a Gaussian fit to the MC reconstructed events
and $B$ is the amount of background extrapolated from the \JpsiPi invariant mass sidebands in the data.
The two sideband regions are defined as being between $5\sigma_{m(\PBc)}$ and $8\sigma_{m(\PBc)}$ of the world-average $\PBc$ mass \cite{PDG},
where $\sigma_{m(\PBc)}$ is the resolution of the signal as determined in simulation.

\begin{figure}[hbtp]
  \begin{center}
    \includegraphics[width=0.45\textwidth]{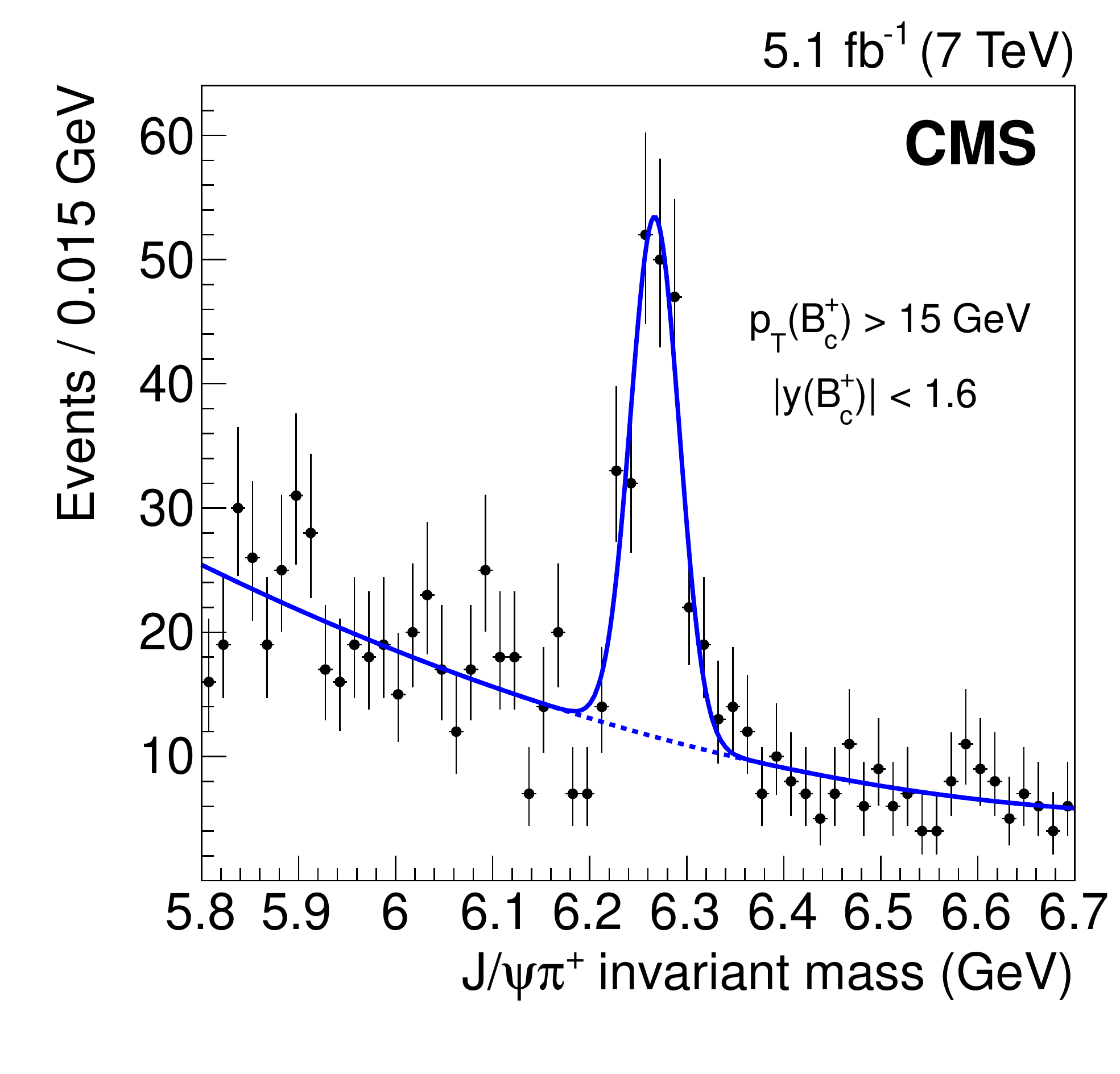}
    \includegraphics[width=0.45\textwidth]{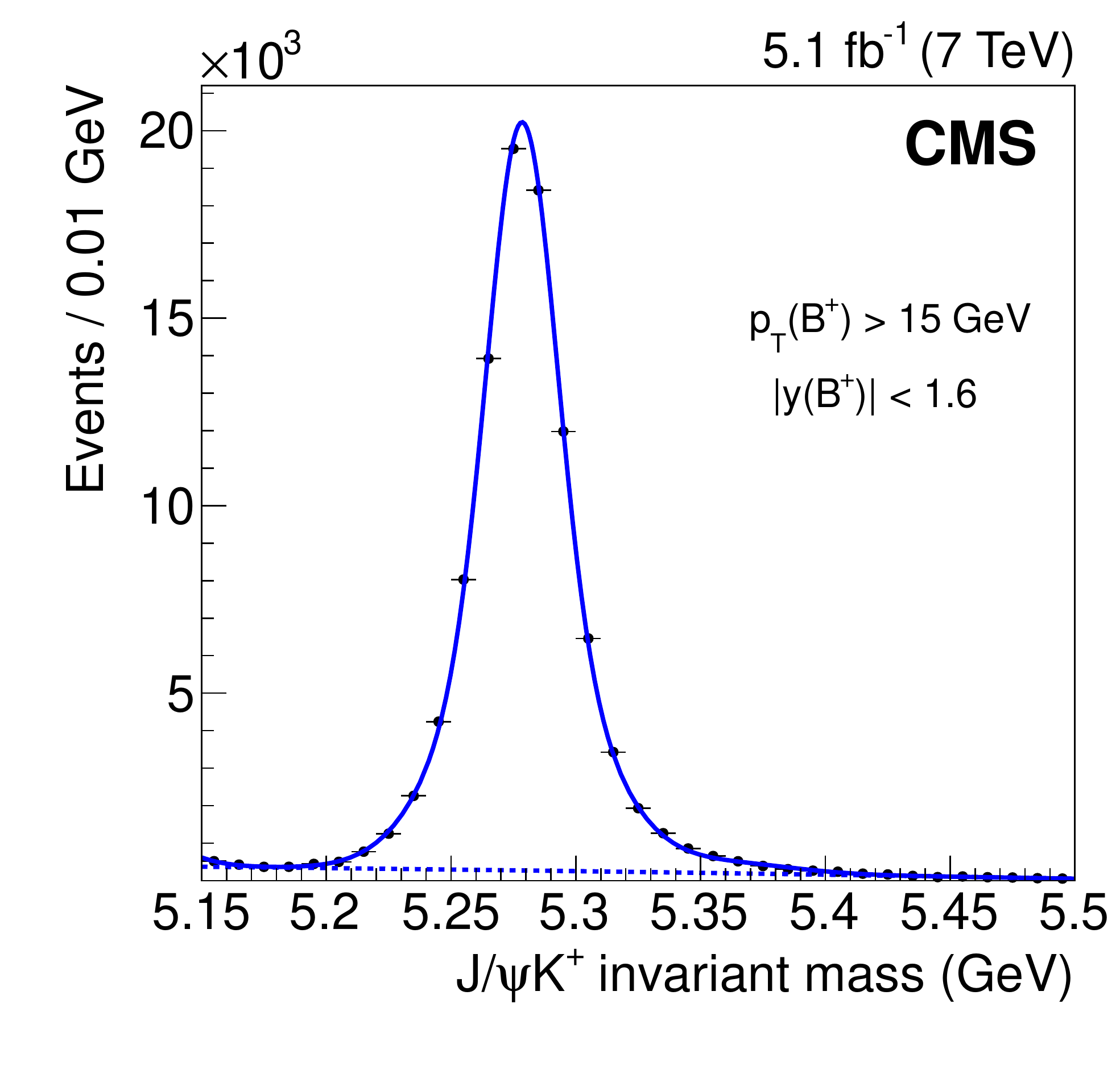}
    \caption{The \JpsiPi (left) and	
     \JpsiK (right) invariant mass distributions. The result of the fit is superimposed. The lines represent the signal-plus-background fit (solid) and the background-only component (dashed).}
    \label{fig:BcSignal}
    \label{fig:BpSignal}
  \end{center}
\end{figure}

The procedure results in the following requirements:
\PBpc vertex probability $>$6\%,
 $\cos\alpha' > 0.9$, where $\alpha'$ is the angle between the candidate \PBpc momentum
vector and the displacement between the beam spot and the decay vertex evaluated in the plane transverse to the beam;
 $\pt(\Pgp)> 2.7$\GeV, and  $\Delta R(\JPsi,\Pgp)<1 $, where
 $\Delta R$ is the distance in the ($\eta,\phi$) plane between the \JPsi and pion momentum vector.
The \BcJpsiPi invariant mass distribution is shown in Fig.~\ref{fig:BcSignal} (left).
The \BJpsiK signal is obtained with the same selections and is shown in Fig.~\ref{fig:BpSignal} (right).
The \BcJpsiPi and the \BJpsiK invariant mass distributions are fit with an unbinned maximum likelihood estimator.
The \PBpc signal is fit with a Gaussian distribution and the background with a second-order Chebyshev polynomial.
The \BcJpsiPi signal has a yield of $176 \pm 19$, a mass of $6.267 \pm 0.003$\GeV, and a resolution of $0.025 \pm 0.003$\GeV (statistical uncertainties only).
Contamination from other \PBpc decay modes in the \BcJpsiPi channel has been investigated.
A possible reflection of the Cabibbo-suppressed \BcJpsiK mode in the \JpsiPi mass spectrum has been modeled  from a simulated sample of \BcJpsiK events and its
contribution constrained using the value of the relative branching fraction to \JpsiPi \cite{lhcb_JpsiK}.
Furthermore, the effect due to a possible undetected $\Pgpz$ from \BcJpsiPiPiO decay has been modeled from a dedicated MC sample.
The partially reconstructed \JpsiPi mass spectrum obtained from the simulated events has been fit with an ARGUS function \cite{ARGUS} convolved with a Gaussian function describing the detector resolution.
The resulting parametrization, added to a linear function, has been used to describe the background on the left of the signal peak in the fit of the \JpsiPi mass spectrum in data.
No significant variation of the \BcJpsiPi signal yield is found.

The $\PBp$ invariant mass distribution is fit with a sum of two Gaussian distributions with a common mean for the signal and a second-order Chebyshev polynomial for the background.
Additional contributions from partially reconstructed $\PBz$ and $\PBp$ decays
are parametrized with functions determined from inclusive $\PBp \to\JPsi X$ and $\PBz \to \JPsi X$ MC samples.

\section{\texorpdfstring{\RatioBcBp}{R(c/u)} measurement}
\label{sec:BcBpRatio}
The ratio \RatioBcBp of the production cross sections times branching fractions is obtained from the relation
\begin{equation}
\label{eq:ratiobp2}
\RatioBcBp = \frac{\sigma(\PBpc) \mathcal{B}(\PBpc \to \JPsi \Pgpp) }
{ \sigma(\PBp)\mathcal{B}(\PBp \to \JPsi \PKp) }=
\frac{Y_{\PBpc \to \JPsi \Pgpp}  } { Y_{\PBp \to \JPsi \PKp} },
\end{equation}
where $Y_{\PBpc \to \JPsi \Pgpp}$ and $Y_{\PBp \to \JPsi \PKp}$ are the signal yields extracted from the efficiency-corrected invariant mass distributions for the \BcJpsiPi
and \BJpsiK channels, respectively, in the kinematic region $\pt >15$\GeV and $\abs{y}< 1.6$. The efficiencies for the two channels are evaluated from MC simulations and  include geometrical acceptance, reconstruction, selection, and trigger effects.

The simulation of the two-body $\PBpc \to \JPsi \Pgpp$ and $\PBp \to \JPsi \PKp$ decays takes into account the spins of the particles.
The efficiencies are evaluated as a function of the $\PBp$  or \PBpc candidate's \pt and computed in \pt bins, whose sizes are determined by the available size of the \BcJpsiPi and \BJpsiK MC samples.
Data are corrected event-by-event according to the candidate's \pt.

Possible systematic uncertainties introduced by different trigger and pileup conditions and analysis selections have been investigated by dividing the data and evaluating
the statistical consistency \cite{PDG} of the independent samples; the resulting systematic uncertainties are found to be insignificant.
Uncertainties from the different signal and background fit functions and fit ranges have been evaluated through a ``fit variant'' approach \cite{FOCUS_variant} and account for
a 5.3\% uncertainty.
The finite size of the MC samples introduces a systematic uncertainty of 2.1\% and the choice of the \pt binning in the efficiency calculation
 an additionl 3.1\%.
The total systematic uncertainty in the ratio is 6.5\%.

Recently, the LHCb Collaboration published a new, more precise \PBpc lifetime measurement \cite{lhcb_newlife}, which is significantly higher than the previous world average \cite{PDG}.
The \BcJpsiPi reconstruction efficiency has a dependence on the \PBpc lifetime. To determine the systematic uncertainty associated with the uncertainty in the \PBpc lifetime, the efficiency is evaluated while changing the \PBpc lifetime in the simulation
to cover the range from the world average minus its one standard deviation uncertainty, to the new LHCb measurement.
The resulting variation in the \RatioBcBp ratio is quoted separately as a lifetime systematic uncertainty $(\sigma(\tau_{\PBc}))$ and is $\pm 10.4\%$.
The different contributions to the systematic uncertainty are listed in Table \ref{tab:ratioBpSyst}.

\begin{table*}[htbH]
\topcaption{Systematic uncertainties in the measurement of \RatioBcBp.}
\label{tab:ratioBpSyst}
\centering
\begin{tabular}{lcc}
\hline
Systematic source				&  \% 	 \\
\hline
Fit variant					    & 5.3		 \\
MC sample size			        & 2.1		 \\
Efficiency binning  			& 3.1	     \\
\hline
Total uncertainty			    & 6.5	      \\
\hline
$\PBc$ lifetime				    & 10.4 	     \\
\hline
\end{tabular}
\end{table*}

The measurement of the ratio, including all the uncertainties, is
\begin{equation}
R_{\cPqc/\cPqu} = [0.48 \pm 0.05\stat \pm 0.03\syst \pm 0.05\,(\tau_{\PBc})]\%.
\end{equation}

 \section{\texorpdfstring{\BcJpsiPiPiPi}{Bc to J/Psi pi pi pi}  decay}

\begin{figure}[hbtp]
  \begin{center}
    \includegraphics[width=0.45\textwidth]{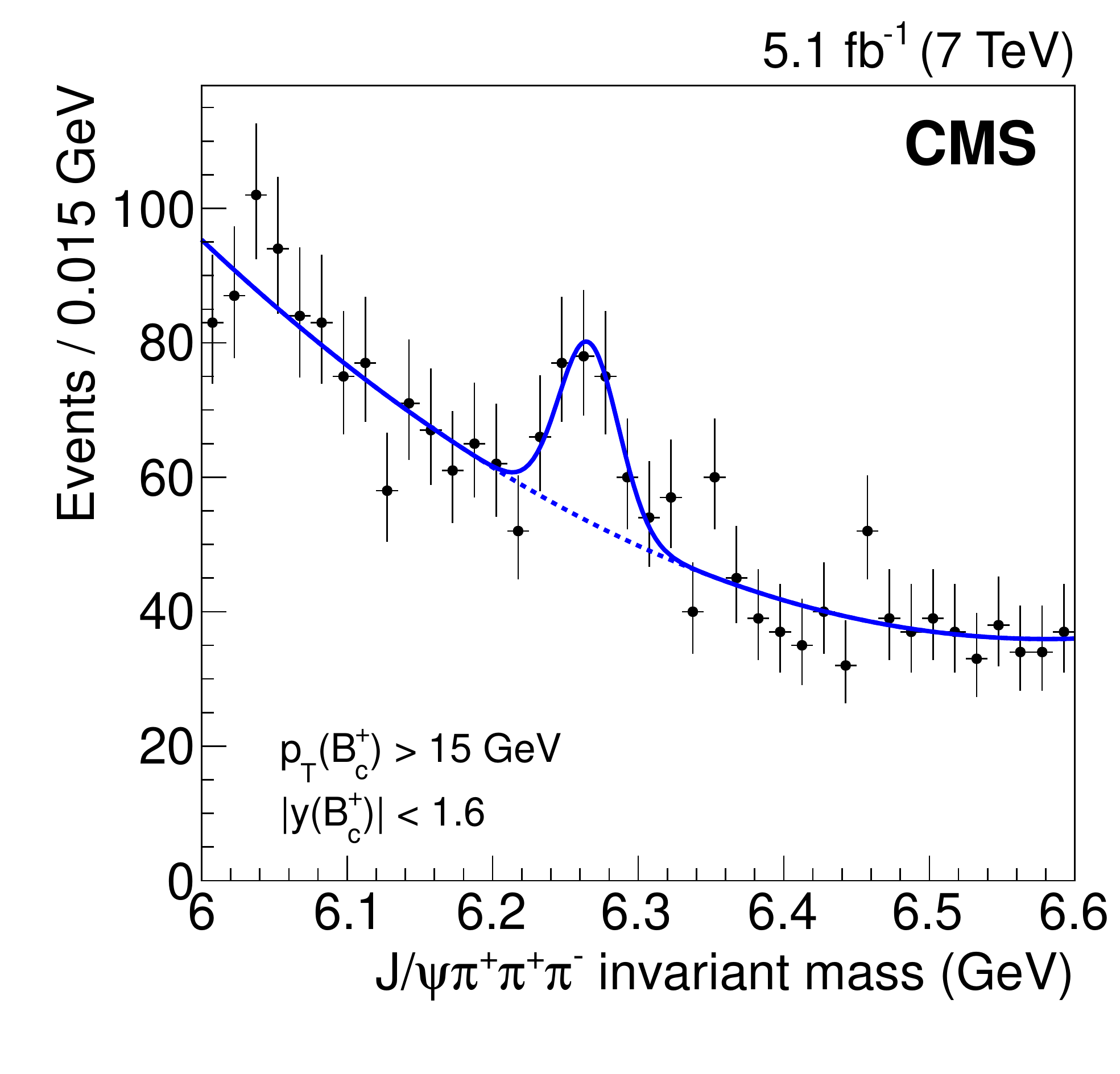}
    \caption{The \JpsiPiPiPi invariant mass distribution. The result of the fit is superimposed; the lines represent the signal-plus-background fit (solid) and the background-only component (dashed).}
    \label{fig:Bc3piSignal}
  \end{center}
\end{figure}

The same figure of merit  $S/\sqrt{(S+B)}$ is maximized in the selection of the \BcJpsiPiPiPi signal in the same kinematic phase space as defined for the \BcJpsiPi decay, i.e.,
$\pt(\PBpc) > 15\GeV$ and  $\abs{y(\PBpc)}< 1.6$.
The optimized selection requirements are:
 $\chi^2$ probability of the five-track kinematic fit $>$20\%;
$\cos\alpha'> 0.99$;
 $\pt(\Pgp{_1})> 2.5$\GeV;
 $\pt(\Pgp{_2})> 1.7$\GeV;
 $\pt(\Pgp{_3})> 0.9$\GeV,
 where the three pions are referred to as $\Pgp_1$, $\Pgp_2$, and $\Pgp_3$ from highest to lowest \pt;
and $\Delta R(\JPsi,\Pgp_S)<0.5$, where $\Pgp_S$ is the sum of the momentum vectors of the three pions.
The resulting \BcJpsiPiPiPi invariant mass distribution is shown in Fig.~\ref{fig:Bc3piSignal}.
A fit is performed with an unbinned maximum likelihood estimator. The signal is parametrized as a Gaussian distribution and the background as a second-order Chebyshev polynomial.
The signal yield is $92 \pm 27$ events and the fitted mass and resolution values are $6.266 \pm 0.006$\GeV and $0.021 \pm 0.006$\GeV, respectively, where the uncertainties are statistical only.
Possible contamination from other \PBpc decay modes in the \BcJpsiPiPiPi channel has been investigated.
No \BcJpsiKKPi decays are observed in the data with the applied selection cuts.
The effect from a possible undetected $\Pgpz$ in the decay \BcJpsiPiPiPiPiO has been modeled with a dedicated MC sample.
The partially reconstructed \JpsiPiPiPi mass spectrum obtained from the simulated events has been fit with an ARGUS function convolved with a Gaussian function describing the detector resolution.
The resulting parametrization, added to a linear polynomial function, has been used to describe the background on the left of the signal
peak in the fit of the \JpsiPiPiPi mass spectrum in data.
No significant variation in the \BcJpsiPiPiPi signal yield is found.

\section{\texorpdfstring{\RatioBcBc}{R(Bc)}  measurement }
\label{sec:BcBcRatio}

\label{sec:effSubmasses}

The ratio \RatioBcBc is defined as
\begin{equation}
\label{eq:ratiobcbc}
R_{\PBc} = \frac{ \mathcal{B}(\PBpc \to \JPsi \Pgpp\Pgpp\Pgpm) }
{\mathcal{B}(\PBpc \to \JPsi \Pgpp)}=
\frac{Y_{\PBpc \to \JPsi \Pgpp \Pgpp \Pgpm} }
{Y_{\PBpc \to \JPsi \Pgpp}},
\end{equation}%
where $Y_{\PBpc \to \JPsi \Pgpp \Pgpp \Pgpm}$ and $Y_{\PBpc \to \JPsi \Pgpp}$ are the signal yields extracted from the
efficiency-corrected invariant mass distributions for the \BcJpsiPiPiPi
and \BcJpsiPi channels, respectively, in the kinematic region $\pt >15$\GeV and $\abs{y}< 1.6$.
Efficiency corrections of the \BcJpsiPiPiPi and \BcJpsiPi data include geometrical acceptance, reconstruction, selection,
and trigger effects. The efficiencies for the two channels are evaluated from MC simulations.

The efficiency for the \BcJpsiPi channel is evaluated as a function of the candidate's \pt, as explained in Section \ref{sec:BcBpRatio}.

\begin{figure}[hbtp]
 \begin{center}
   \subfloat{\includegraphics[width=0.35\textwidth]{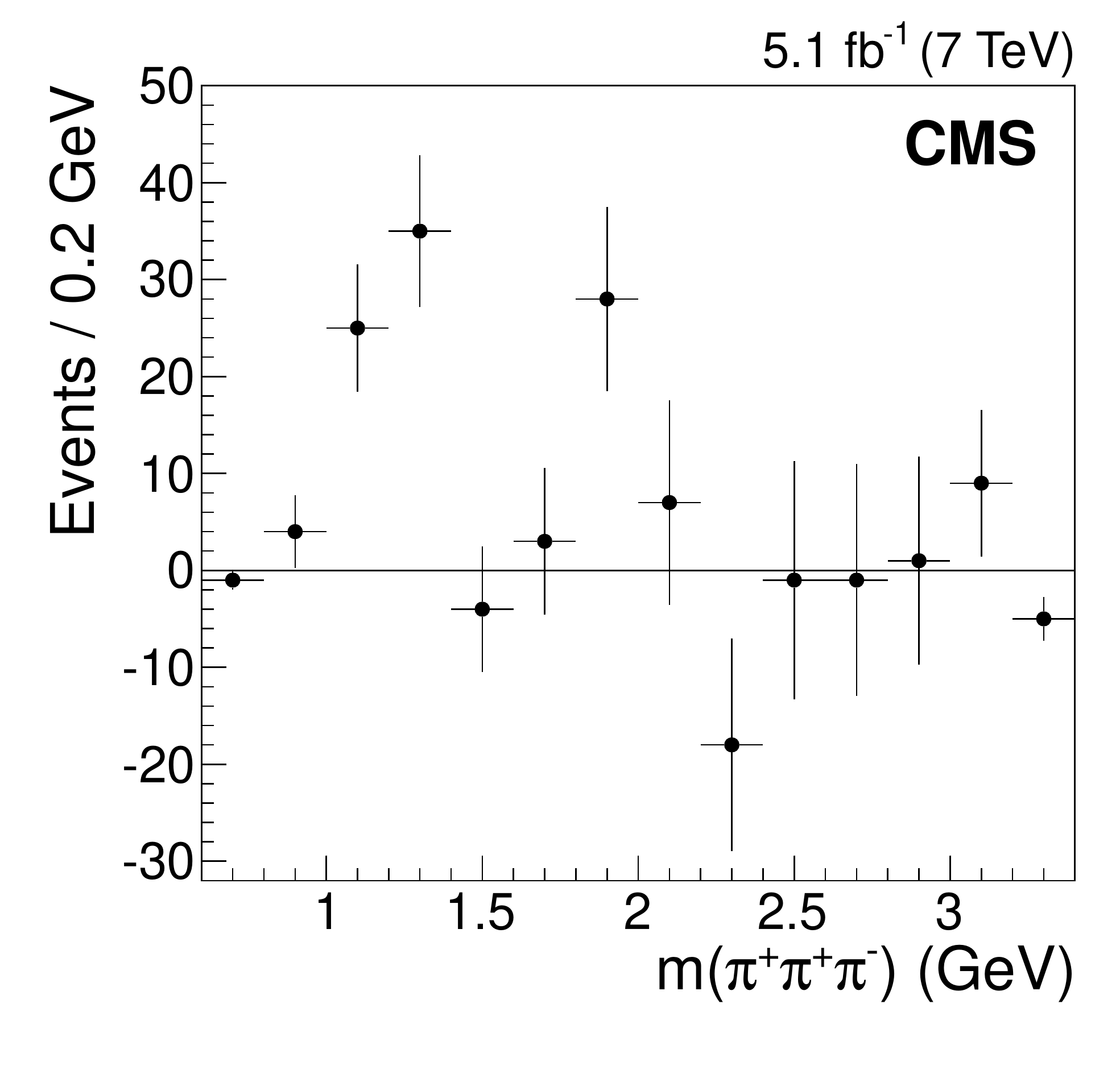}}
   \subfloat{\includegraphics[width=0.35\textwidth]{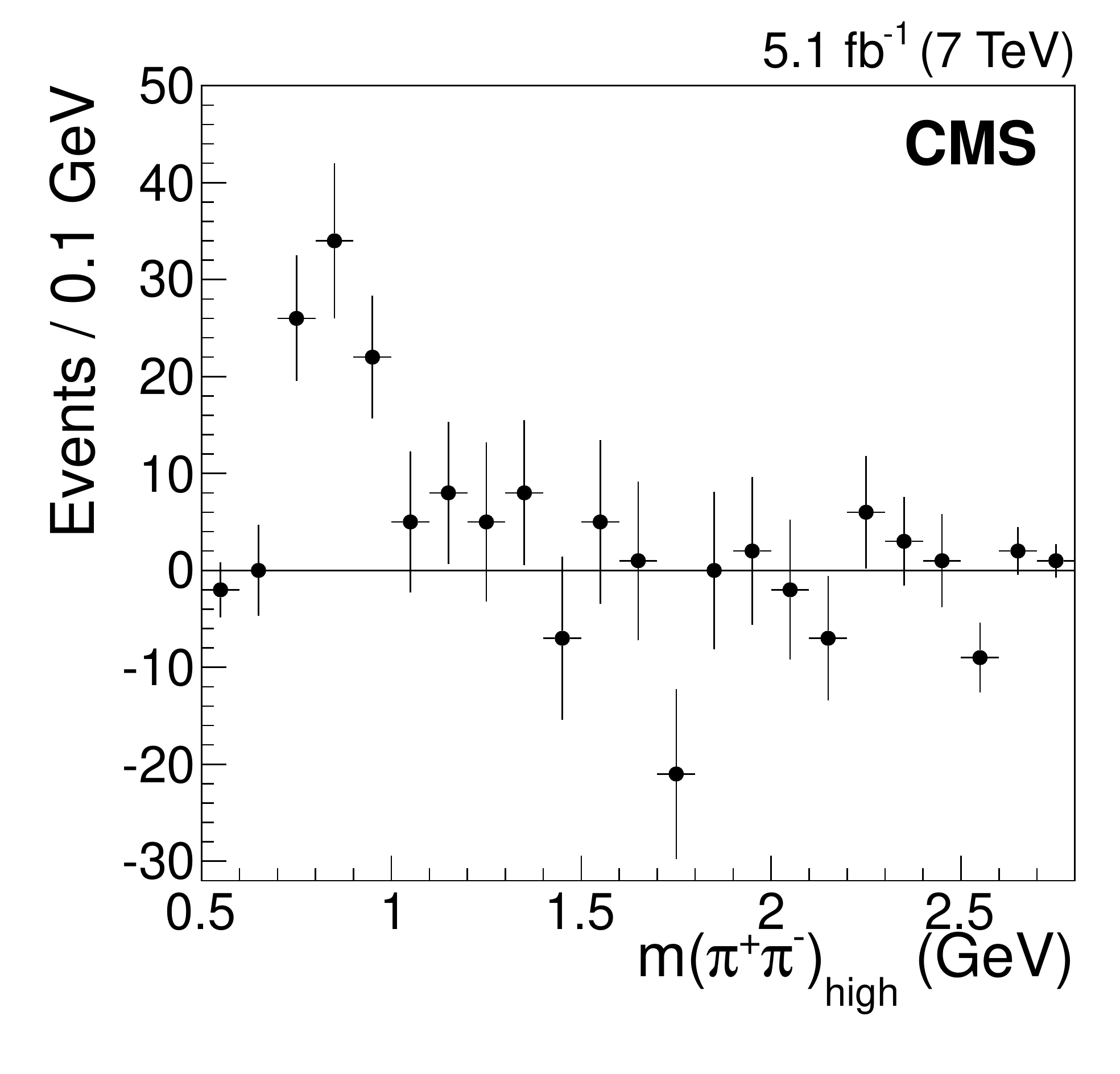}}
   \subfloat{\includegraphics[width=0.35\textwidth]{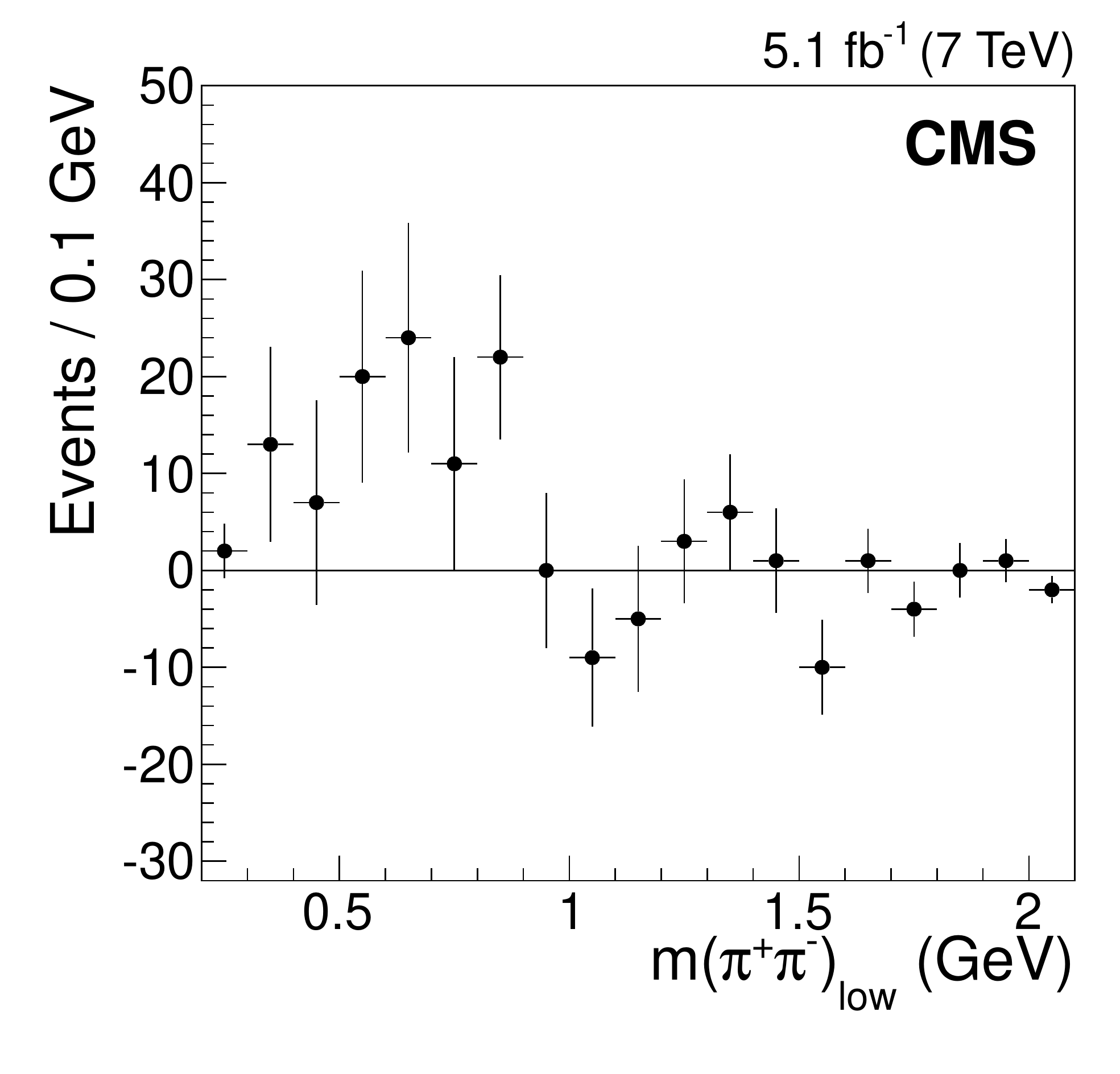}}
   \caption{Background-subtracted invariant mass projections for $\Pgpp\Pgpp\Pgpm$ (left), $(\Pgpp\Pgpm)_{\text{high}}$ (center), and
   $(\Pgpp\Pgpm)_{\text{low}}$ (right) from the \BcJpsiPiPiPi candidate events.
   Since two same-sign pions are present in the final state, we indicate with $(\Pgpp\Pgpm)_{\text{low}}$ the $\Pgpp\Pgpm$ pair with the lower invariant mass and with $(\Pgpp\Pgpm)_{\text{high}}$ the higher invariant mass combination. }
   \label{fig:substructuresProjection}
 \end{center}
\end{figure}

The \BcJpsiPiPiPi decay can involve intermediate resonant states; indeed, the $\Pgpp\Pgpp\Pgpm$ and $\Pgpp\Pgpm$ invariant mass projections from data show evidence for the presence of
$\Pai$ and $\Pgr$ in the decay (Fig.~\ref{fig:substructuresProjection}).
No hint of either $\Pgy (\to  \JPsi \Pgpp \Pgpm)$ or $\PX(3872)(\to  \JPsi \Pgpp \Pgpm)$ is detected in the $\Pgmp\Pgmm\Pgpp\Pgpm$ mass projections.
The quantitative determination of the resonant contributions and their interferences in the decay requires a sophisticated amplitude analysis which is not feasible with the available amount of data.
However, the reconstruction efficiency for this five-body decay could be affected by the decay dynamics; thus, a model-independent efficiency treatment is needed.

A five-body decay of a spinless particle can be fully described in its center-of-mass frame by eight independent mass combinations of the type $m^2_{ij}$ ($i \neq j$), where $m^2_{ij}$ is the squared invariant mass of the pair of
particles $i$ and $j$ in the final state (Dalitz plot representation).
 In the present case, the additional \JPsi mass constraint reduces the number of independent $m^2_{ij}$ to seven. The following seven mass combinations have been chosen:
$x=m^2(\Pgmp\Pgpp)_{\text{low}}$, $y=m^2({\Pgpp\Pgpm})_{\text{high}}$, $z=m^2({\Pgmp\Pgpm})$, $w=m^2({\Pgpp\Pgpp})$, $r=m^2({\Pgmm\Pgpp})_{\text{low}}$,
$t=m^2({\Pgmm\Pgpp})_{\text{high}}$, and $v=m^2({\Pgmm\Pgpm})$;
the ``low" and ``high" subscripts refer to the lower and higher invariant mass combination where a $\Pgpp$ is involved.
A \BcJpsiPiPiPi nonresonant MC has been produced to access all the phase-space configurations.
The efficiency is parametrized as a linear function of these seven mass combinations:
\begin{equation}
\label{eq:effpol}
\epsilon=\abs{p_0 + p_1\cdot x + p_2 \cdot y + p_3 \cdot z + p_4 \cdot w +  p_5 \cdot r + p_6 \cdot t + p_7 \cdot v},
\end{equation}
where $p_i$ are free parameters to be determined via an unbinned maximum likelihood fit to the generated events in the seven-dimensional space using a binomial probability density function.
The absolute value is required to prevent the function from assuming negative values.
The resulting efficiency function is used to weight the data event-by-event.
The data efficiency-corrected invariant mass for the \BcJpsiPi and \BcJpsiPiPiPi channels is fit
with a function consisting of a Gaussian distribution for the signal and a second-order Chebyshev polynomial for the background.
An unbinned maximum likelihood estimator is used to extract the $Y_{\PBpc \to \JPsi \Pgpp}$ and $Y_{\PBpc \to \JPsi \Pgpp \Pgpp \Pgpm}$ yields.
The resulting measurement of the branching fraction ratio is $\RatioBcBc = 2.55 \pm 0.80$,
where the uncertainty is statistical only.

The same sources of systematic uncertainties considered in Section \ref{sec:BcBpRatio} have been evaluated for this measurement as well; the biggest uncertainty comes from the fit variant and accounts for a 9.4 \% effect.
An additional contribution is considered coming from the choice of the seven-dimensional efficiency parametrization of Eq. (\ref{eq:effpol}).
To estimate this contribution, data are alternatively weighted according to the efficiency distribution directly obtained from the MC samples, binned in the seven two-body submasses.
The difference between the ratio measured using the binned efficiency distribution and
the function in Eq.~(\ref{eq:effpol})
is taken as a systematic uncertainty (1.0\%).
In the evaluation of \RatioBcBc, two different multiplicity final states are compared. Assuming a tracking efficiency uncertainty for each pion track
of 3.9\% \cite{CMS-PAS-TRK-10-002}, a global 7.8\% uncertainty is included in the final systematic evaluation.
The total systematic uncertainty in the ratio, obtained by adding all the contributions in quadrature, is 13.1\%.

The systematic uncertainty from the \PBpc lifetime uncertainty is quoted separately in the ratio and is $(\sigma(\tau_{\PBc}))$ =$^{+1.6}_{-0.4}$\%.
The sources of systematic uncertainty are summarized in Table \ref{tab:ratioBcSyst}.

\begin{table*}[htbH]
\topcaption{Systematic uncertainties in the measurement of \RatioBcBc}.
\label{tab:ratioBcSyst}
\centering
\begin{tabular}{l c}
\hline
Systematic source   					& \%		                   \\
\hline
Fit variant						& 9.4		 		   \\
MC sample size			    		        & 4.1				   \\
Efficiency fit function			                & 1.0				   \\
Efficiency binning 				        & 1.9				   \\
Tracking efficiency				        & 7.8				   \\
\hline
Total uncertainty			                & 13.1			           \\
\hline
Lifetime				      		&$^{+1.6}_{-0.4}$                  \\
\hline
\end{tabular}
\end{table*}

The resulting ratio, including all uncertainties, is
\begin{equation}
R_{\PBc} = 2.55 \pm 0.80\stat \pm 0.33\syst \,^{+ 0.04}_{-0.01} \,(\tau_{\PBc}).
\end{equation}

\section{Summary}

A measurement of the ratio of the cross sections times branching fractions for \BcJpsiPi and \BJpsiK has been presented based on pp collision data at a center-of-mass energy of 7\TeV collected by the CMS experiment and corresponding to an integrated
luminosity of 5.1\fbinv.
The analysis, performed for \PBpc and $\PBp$ mesons with $\pt>
 15\GeV$ and in the central rapidity region $\abs{y}< 1.6$, gives a measured ratio of
\begin{equation}
R_{\cPqc/\cPqu} = [0.48 \pm 0.05\stat \pm 0.03\syst \pm 0.05\,(\tau_{\PBc})]\%.
\end{equation}

A similar measurement from LHCb in the kinematic region $\pt>
 4\GeV$, $ 2.5<\eta <4.5$ gives $[0.68 \pm 0.10\stat \pm 0.03\syst \pm 0.05\,(\tau_{\PBc})]\%$ \cite{lhcbR}.
The two measurements, performed in different kinematic regions, are expected to differ because of the softer \pt distribution of the \PBpc with respect to that of the $\PBp$,
implying a lower value of the ratio at higher \pt.
The measurements are consistent with this expectation.
Measurements of the production cross section times branching fraction for $\PBpc \to \JPsi\ell^+ \nu$ relative to that for
\BJpsiK are also available from the CDF experiment \cite{CDF_sl} in the kinematic region $\pt
> 4\GeV$  and $\abs{y}< 1$.
With the present \PBpc (\pt,$\abs{y}$) coverage, these experimental results can give guidance to improve the theoretical
calculations still affected by large uncertainties and constrain the various \PBpc production models.

The ratio of the \BcJpsiPiPiPi and \BcJpsiPi branching fractions has been measured to be
\begin{equation}
R_{\PBc} = 2.55 \pm 0.80\stat \pm 0.33\syst^{+ 0.04}_{-0.01}\,(\tau_{\PBc}),
\end{equation}

which is in good agreement with the result from the LHCb experiment, $2.41 \pm 0.30\stat \pm 0.33\syst$ \cite{lhcb3to1}, and represents its first confirmation.
This measurement can be compared with the theoretical predictions, which assume factorization into $\PBpc \to \JPsi \PWp^*$ and $\PWp^* \to n\Pgpp$ ($n = 1$, 2, 3, 4).
In particular, Ref.~\cite{production_disagreement} predicts 1.5 for the ratio, whereas Ref.~\cite{thratio} predicts three different values,
1.9, 2.0, and 2.3, depending on the chosen set of \PBpc meson form factors.
More precise measurements are needed to determine if one of the predictions is favored by the data.
The model-independent method implemented for the efficiency evaluation of the five-body final state can be considered in future high-statistics analyses to reduce systematic uncertainties associated with the unknown multibody decay dynamics.

\begin{acknowledgments}
We congratulate our colleagues in the CERN accelerator departments for the excellent performance of the LHC and thank the technical and administrative staffs at CERN and at other CMS institutes for their contributions to the success of the CMS effort. In addition, we gratefully acknowledge the computing centers and personnel of the Worldwide LHC Computing Grid for delivering so effectively the computing infrastructure essential to our analyses. Finally, we acknowledge the enduring support for the construction and operation of the LHC and the CMS detector provided by the following funding agencies: BMWFW and FWF (Austria); FNRS and FWO (Belgium); CNPq, CAPES, FAPERJ, and FAPESP (Brazil); MES (Bulgaria); CERN; CAS, MoST, and NSFC (China); COLCIENCIAS (Colombia); MSES and CSF (Croatia); RPF (Cyprus); MoER, ERC IUT and ERDF (Estonia); Academy of Finland, MEC, and HIP (Finland); CEA and CNRS/IN2P3 (France); BMBF, DFG, and HGF (Germany); GSRT (Greece); OTKA and NIH (Hungary); DAE and DST (India); IPM (Iran); SFI (Ireland); INFN (Italy); NRF and WCU (Republic of Korea); LAS (Lithuania); MOE and UM (Malaysia); CINVESTAV, CONACYT, SEP, and UASLP-FAI (Mexico); MBIE (New Zealand); PAEC (Pakistan); MSHE and NSC (Poland); FCT (Portugal); JINR (Dubna); MON, RosAtom, RAS and RFBR (Russia); MESTD (Serbia); SEIDI and CPAN (Spain); Swiss Funding Agencies (Switzerland); MST (Taipei); ThEPCenter, IPST, STAR and NSTDA (Thailand); TUBITAK and TAEK (Turkey); NASU and SFFR (Ukraine); STFC (United Kingdom); DOE and NSF (USA).

Individuals have received support from the Marie-Curie programme and the European Research Council and EPLANET (European Union); the Leventis Foundation; the A. P. Sloan Foundation; the Alexander von Humboldt Foundation; the Belgian Federal Science Policy Office; the Fonds pour la Formation \`a la Recherche dans l'Industrie et dans l'Agriculture (FRIA-Belgium); the Agentschap voor Innovatie door Wetenschap en Technologie (IWT-Belgium); the Ministry of Education, Youth and Sports (MEYS) of the Czech Republic; the Council of Science and Industrial Research, India; the HOMING PLUS programme of Foundation for Polish Science, cofinanced from European Union, Regional Development Fund; the Compagnia di San Paolo (Torino); the Consorzio per la Fisica (Trieste); MIUR project 20108T4XTM (Italy); the Thalis and Aristeia programmes cofinanced by EU-ESF and the Greek NSRF; and the National Priorities Research Program by Qatar National Research Fund.
\end{acknowledgments}
\bibliography{auto_generated}   % will be created by the tdr script.

\cleardoublepage \appendix\section{The CMS Collaboration \label{app:collab}}\begin{sloppypar}\hyphenpenalty=5000\widowpenalty=500\clubpenalty=5000\textbf{Yerevan Physics Institute,  Yerevan,  Armenia}\\*[0pt]
V.~Khachatryan, A.M.~Sirunyan, A.~Tumasyan
\vskip\cmsinstskip
\textbf{Institut f\"{u}r Hochenergiephysik der OeAW,  Wien,  Austria}\\*[0pt]
W.~Adam, T.~Bergauer, M.~Dragicevic, J.~Er\"{o}, C.~Fabjan\cmsAuthorMark{1}, M.~Friedl, R.~Fr\"{u}hwirth\cmsAuthorMark{1}, V.M.~Ghete, C.~Hartl, N.~H\"{o}rmann, J.~Hrubec, M.~Jeitler\cmsAuthorMark{1}, W.~Kiesenhofer, V.~Kn\"{u}nz, M.~Krammer\cmsAuthorMark{1}, I.~Kr\"{a}tschmer, D.~Liko, I.~Mikulec, D.~Rabady\cmsAuthorMark{2}, B.~Rahbaran, H.~Rohringer, R.~Sch\"{o}fbeck, J.~Strauss, A.~Taurok, W.~Treberer-Treberspurg, W.~Waltenberger, C.-E.~Wulz\cmsAuthorMark{1}
% ...
\vskip\cmsinstskip
% ...
1:~~Also at Vienna University of Technology, Vienna, Austria\\
2:~~Also at CERN, European Organization for Nuclear Research, Geneva, Switzerland\\
% ...

\end{sloppypar}
\end{document}